\documentclass[aps,prl,twocolumn,superscriptaddress,footinbib,showpacs,amsfonts,amssymb,amsmath]{revtex4}
\usepackage{graphicx,bm}
\begin{document}
\title{Spin-lattice order in frustrated ZnCr$_{\textbf{2}}$O$_{\textbf{4}}$}

\author{S. Ji}
\affiliation{Department of Physics, University of Virginia, Charlottesville, VA 22904-4714, USA}
\affiliation{NIST Center for Neutron Research, National Institute of Standards and Technology, Gaithersburg, MD 20899, USA}
\author{S.-H. Lee}
\email{shlee@virginia.edu}
\affiliation{Department of Physics, University of Virginia, Charlottesville, VA 22904-4714, USA}
\author{C. Broholm}
\affiliation{Department of Physics and Astronomy, Johns Hopkins University, Baltimore, MD 21218, USA}
\author{T. Y. Koo}
\affiliation{Pohang Accelerator Laboratory, Pohang University of Science and Technology, Pohang, 790-784, South Korea}
\author{W. Ratcliff}
\affiliation{NIST Center for Neutron Research, National Institute of Standards and Technology, Gaithersburg, MD 20899, USA}
\author{S. -W. Cheong}
\affiliation{Department of Physics and Astronomy, Rutgers University, Piscataway, NJ 08854, USA}
\author{P. Zschack}
\affiliation{Frederick-Seitz Materials Research Lab, University of Illinois at Urbana-Champaign, IL 61801, USA}

\date{\today}

\begin{abstract}
Using synchrotron X-rays and neutron diffraction we disentangle spin-lattice order in highly frustrated ZnCr$_2$O$_4$ where magnetic chromium ions occupy the vertices of regular tetrahedra. Upon cooling below 12.5 K the quandary of anti-aligning spins surrounding the triangular faces of tetrahedra is resolved by establishing weak interactions on each triangle through an intricate lattice distortion. 
The resulting spin order is however, not simply a N\'{e}el state on strong bonds. 
A complex co-planar spin structure indicates that antisymmetric and/or further neighbor exchange interactions also play a role as ZnCr$_2$O$_4$ resolves conflicting magnetic interactions.

\end{abstract}

\pacs{}
\maketitle

While tetrahedral atomic clusters are a natural consequence of close packing, they are particularly inconvenient for antiferromagnetically interacting spins. 
This is because no spin configuration can  simultaneously satisfy all six antiferromagnetic interactions amongst spins on the vertices of a tetrahedron \cite{anderson73, moessner06, bramwell01, moessner98, canals98}.
The consequence of such ``geometrical frustration'' is deep suppression of magnetic order and a range of temperatures where spins remain fluctuating despite interactions that far exceed thermal energies \cite{shlee02, mirebeau02}. 
Indeed for spins on a lattice of corner-sharing tetrahedra, it appears there is no conventional order in the quantum limit ($S$ = 1/2, $T$ = 0) \cite{canals98}. 
Because they entail higher energy spin configurations, geometrically frustrating lattices however typically do not survive in the low temperature limit. 
Instead a compromise between spin and lattice energy is reached through a first order phase transition that freezes the spin liquid and distorts the lattice \cite{freltoft88, shlee00, jhchung05, ueda06, rudolf07, yokaichiya09}. 
Such phase transitions challenge conventional theories of magnetism because they involve strongly correlated spins and the collapse of the rigid lattice approximation \cite{yamashita00,tchernyshyov02,khomskii05}. 

\begin{figure}
\includegraphics[width=0.48\textwidth]{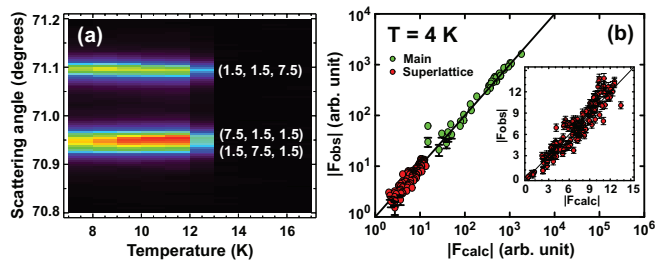}
\centering
\caption{(Color online)
(a) Temperature dependence of the peak intensity of the superlattice Bragg reflection at (7.5, 1.5, 1.5)$_\textrm{c}$. 
Superlattice peak appears below the cubic-to-tetragonal phase transition at $T_{\textrm{N}}$ = 12.5 K. 
(b) Measured ($y$-axis) and calculated ($x$-axis) values for the absolute nuclear structure factors of 44 main (green symbols) and $\sim$ 140 nonequivalent superlattice (red symbols) Bragg reflections on a logarithmic scale. 
The data were taken at 4 K. The inset shows the superlattice data on a linear scale.\label{fig1}
}
\end{figure}

A case in point is ZnCr$_2$O$_4$. At room temperature, it  has a cubic $Fd\bar{3}m$ crystal structure where Cr$^{3+}$ ($S$ = 3/2) ions form a network of corner-sharing tetrahedra \cite{shlee00}. 
The Curie-Weiss temperature is -390 K indicating strong antiferromagnetic frustration, yet chromium spins remain in a cooperative paramagnetic phase down to $T_{\textrm{C}}$ = 12.5 K \cite{shlee02, shlee00}. 
There, a first order phase transition from a cubic paramagnet to a tetragonal antiferromagnet signals the end of distinct spin and lattice degrees of freedom. 
Tetragonal strain energy alone does not account for the difference between magnetic energy gain and overall latent heat and this was a first indication of a more comprehensive rearrangement of the lattice \cite{shlee00}. 
Subsequently X-ray superlattice peaks were detected at $(\frac{1}{2}\frac{1}{2}\frac{1}{2})_{\textrm{c}}$ type reflections (see Fig. \ref{fig1} (a)) \cite{shlee07}. 
This indicates that below $T_{\textrm{N}}$ the tetragonal lattice has $I\bar{4}m2$ symmetry and a $\sqrt{2}\times\sqrt{2}\times 2$ chemical unit cell \cite{note1}. Theoretical efforts to understand the nature of the phase transition have focused on magneto-elastic couplings that involve symmetric isotropic nearest neighbor (NN) exchange interactions \cite{yamashita00,tchernyshyov02,khomskii05}. 

Here we report a combined synchrotron X-ray and magnetic neutron diffraction study to determine the low $T$ spin-lattice order in ZnCr$_2$O$_4$. 
The principal findings are as follows:  
(1) The $I\bar{4}m2$ tetragonal crystal structure features a non-uniform pattern of exchange interactions in which tetrahedra have either two strong and four weak bonds or four strong and two weak bonds. 
Considering strong bonds only, the lattice is reorganized into four disjoint sublattices that no longer frustrate nearest neighbor isotropic exchange interactions. 
(2) Every tetrahedron has two pairs of antiparallel spins, forming a non-collinear structure with spins in the tetragonal basal plane. 
The antiferromagnetic spin-pairs however, do not completely match the pattern of the strong NN bonds, indicating that a theoretical account of the phase transition will require going beyond nearest neighbor isotropic exchange and magneto-elasticity.

A 20 mg single crystal and a 30 g powder sample of ZnCr$_2$O$_4$ were used for the synchrotron X-ray and powder neutron diffraction experiments, respectively. Most X-ray measurements including the superlattice reflections shown in Fig. 1 were carried out at the 33BM-C beamline at the Advanced Photon Source of Argonne National Laboratory, while some integer reflections were re-checked to confirm the symmetry of the tetragonal crystal structure at the Pohang Accelerator Laboratory. The neutron powder diffraction measurements were performed at the BT1 neutron diffractometer at the National Institute of Standards and Technology Center for Neutron Research.

\begin{table*}
  \caption{The Cr positions in the tetragonal phase determined from Rietveld refinement of x-ray single crystal diffraction data shown in Fig. \ref{fig1} (b). Displacements of the Cr ions from cubic positions are denoted by $d\bm{r}$ = [$dx$, $dy$, $dz$] in direct lattice coordinates. \label{table1} }
\begin{ruledtabular}
\begin{tabular}{lrrrrrr}
 & x & y & z & $dx$ (10$^{-4}$) & $dy$ (10$^{-4}$) & $dz$ (10$^{-4}$) \\ \hline 
Cr$_\textrm{I}$ (8$i$)  & 0.125 + $dx$ & 0 & 0.1875 + $dz$ &  8.94566(202) & 0 & 13.7973(385) \\ 
Cr$_\textrm{II}$ (8$i$) & 0.375 + $dx$ & 0 & 0.1875 + $dz$ &  6.35547(168) & 0 & -3.35179(70) \\ 
Cr$_\textrm{III}$ (8$i$)& 0.375 + $dx$ & 0 & 0.6875 + $dz$ & -6.35547(168) & 0 &  3.35179(70) \\ 
Cr$_\textrm{IV}$ (8$i$) & 0.125 + $dx$ & 0 & 0.6875 + $dz$ & -8.94566(202) & 0 &-13.7973(385) \\ 
Cr$_\textrm{V}$ (16$j$) & 0.375 + $dx$ & 0.25 + $dy$ & 0.4375 + $dz$ & 5.47996(127) & 5.41282(133) & 2.25274(56) \\ 
Cr$_\textrm{VI}$(16$j$) & 0.875 + $dx$ & 0.75 + $dy$ & 0.4375 + $dz$ &-5.47996(127) &-5.41282(133) &-2.25274(56)  
\end{tabular} 
\end{ruledtabular}
\end{table*}

\begin{figure*}
\includegraphics[width=0.90\textwidth]{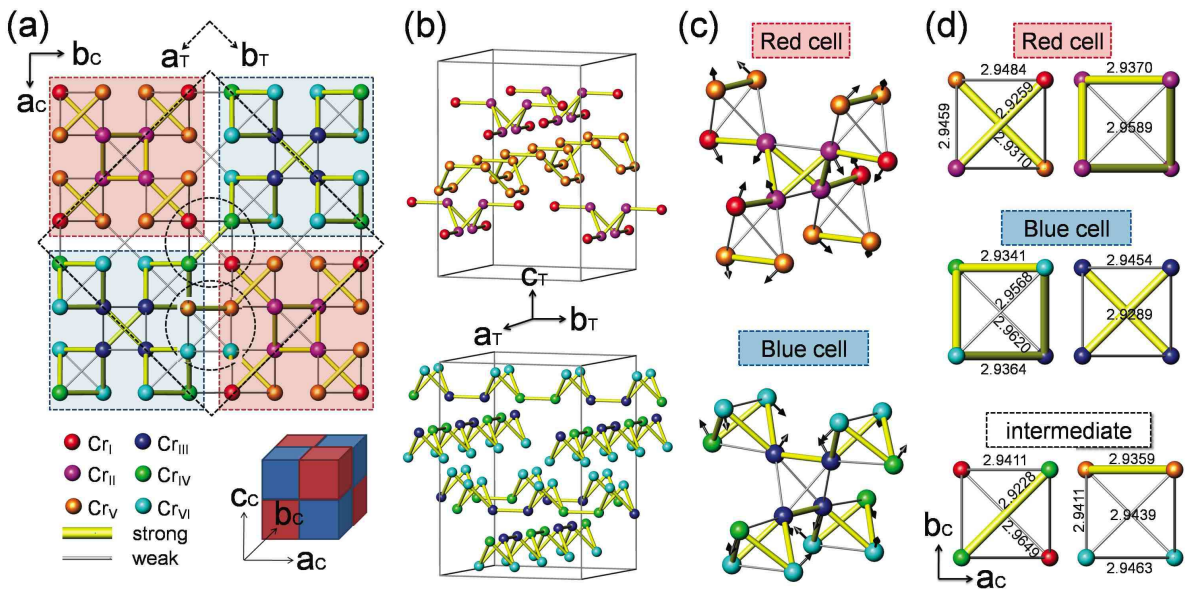}
\centering
\caption{(Color online)
(a) $a$-$b$ projection of the Cr sites in ZnCr$_2$O$_4$. Spheres in different colors represent different Cr sites in the tetragonal phase: Cr$_{\textrm{I}}$ (red), Cr$_{\textrm{II}}$ (violet), Cr$_{\textrm{III}}$ (dark blue), Cr$_{\textrm{IV}}$ (green), Cr$_{\textrm{V}}$ (orange), and Cr$_{\textrm{VI}}$ (light blue). 
The yellow bar and the grey line between Cr ions are short (strong) and long (weak) bonds, respectively. 
The red and blue shaded squares are the cubic unit cells that have different pattern of distortions as shown in Fig. 2 (c), expanding the tetragonal unit cell by $\sqrt{2}\times\sqrt{2}\times 2$ compared to the cubic unit cell. 
(b) Decoupled sublattices that emerge when only the strong bonds are considered. 
(c) Detail of different patterns of distortion in the red and blue cells of a unit cell. 
Black arrows indicate Cr distortions in the tetragonal phase. 
The magnitudes of the distortions are listed in Table \ref{table1}. 
(d) Distorted tetrahedra in the tetragonal phase. The numbers indicate distances in \AA~between Cr ions. 
\label{fig2}}
\end{figure*}

To determine the tetragonal structure in detail, we measured the X-ray integrated intensity of $\sim$ 140 different $(\frac{1}{2}\frac{1}{2}\frac{1}{2})_{\textrm{c}}$ type superlattice reflections through rocking scans. 
Fitting the data within $I\bar{4}m2$, we found sensitivity to Cr positions but not Zn and O positions so changes in the latter positions were ignored in the analysis. 
This is consistent with magneto-elastically driven displacements, $d\bm{r}_i$, of the magnetic Cr ions. 
At low $T$, these Cr ions occupy six crystallographically distinct sites: four $8i$ sites and two $16j$ sites (see Table \ref{table1} and Fig. \ref{fig2} (a)). 
Fourteen parameters are needed to describe the displacements $(d\bm{r}_i = [dx_i, dy_i, dz_i])$ of the six Cr sites. 
However, we did not detect Bragg peaks with indexes such as (4$n$+2 0 0)$_c$ and (2$n$+1 0 0)$_c$, which indicates that $d\bm{r}_4$ = $-d\bm{r}_1$, $d\bm{r}_3$ = $-d\bm{r}_2$, and $d\bm{r}_6$ = $-d\bm{r}_5$. This reduces the number of independent parameters to seven. 
An excellent fit of the main and superlattice peak intensities (green and red symbols in Fig. \ref{fig1} (b), respectively) is obtained within $I\bar{4}m2$ with the displacements listed in Table I and illustrated as black arrows in Fig. \ref{fig2} (c). 

Recall that in the cubic phase each tetrahedron contains four Cr$^{3+}$ ions with six equivalent bonds that lead to geometrical frustration. 
In the tetragonal phase, however, the displacements distinguish Cr$^{3+}$ pairs and lead to 19 different bond lengths, varying from 2.9228 \AA (Cr$_{\textrm{IV}}$-Cr$_{\textrm{IV}}$ bonding in the ab-plane) to 2.9649 \AA (Cr$_{\textrm{I}}$-Cr$_{\textrm{I}}$ bonding in the ab-plane) (see Fig. \ref{fig2} (d)). 
Two aspects of the tetragonal structure should be noted: 
(1) Each tetrahedron has either two strong and four weak bonds (type I) or four strong and two weak bonds (type II) or one strong and five weak bonds (type III). 
This breaks the frustrating isosceles triangular motif of the tetrahedra. 
(2) Considering strong bonds only, the entire pyrochlore lattice is divided into four different sublattices connected by weak bonds only. 
Fig. \ref{fig2} (b) shows that Cr$_{\textrm{I}}$ and Cr$_{\textrm{II}}$ sites form four-legged buckled squares while Cr$_V$ sites form separate buckled octagons - both finite sized spin clusters.  
Cr$_{\textrm{III}}$, Cr$_{\textrm{IV}}$ and Cr$_{\textrm{VI}}$ sites on the other hand form chains of buckled squares along cubic [1 1 0] and [1 $\bar{1}$ 0] directions. 
None of these structures feature a triangular motif so the lattice distortion evidently accomplishes the objective of relieving frustration when only the strong bonds are considered. 

\begin{figure}
\includegraphics[width=0.48\textwidth]{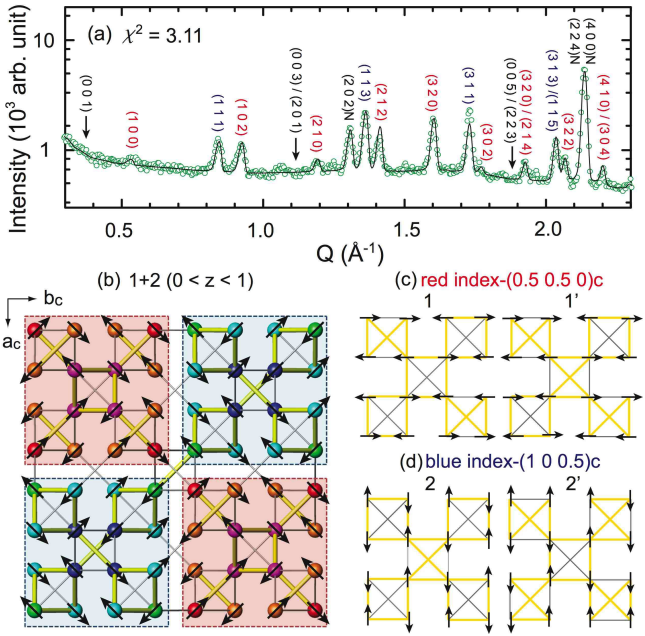}
\centering
\caption{(Color online)
(a) Neutron powder diffraction pattern as a function of wave vector $\bm{Q}$. 
Symbols indicate observed intensities, and the line is the calculated intensity based on the  tetragonal lattice structure and the spin structure indicated in Fig. 3 (b). 
The reflection indexes in the tetragonal notation written in red and in blue belong to $\bm{k}_1$ = ($\frac12$ $\frac12$ 0)$_\textrm{c}$ = (1 0 0)$_\textrm{t}$ and $\bm{k}_2$ = (1 0 $\frac12$)$_\textrm{c}$ = (1 1 1)$_\textrm{t}$, respectively. 
The black arrows represent positions of three (0 0 $\frac12$)$_\textrm{c}$ = (0 0 1)$_\textrm{t}$ type reflections, showing the absence of such magnetic scattering. 
(b) One of 32 coplanar non-collinear spin configurations that give the same best fit to the data with the ordered moment per each Cr$^{3+}$ ion . Thick yellow (thin grey) bars represent strong (weak) bonds. 
When the spin configurations are decomposed into $a$- and $b$-components, one component forms a collinear spin structure with $\bm{k}_1$ = ($\frac12$ $\frac12$ 0)$_\textrm{c}$ while the other forms a collinear spin structure with $\bm{k}_2$ = (1 0 $\frac12$)$_\textrm{c}$. Two patterns of antiparallel spins formed for (c) $\bm{k}_1$ and (d) $\bm{k}_2$. 
Yellow lines in (c) and (d) simply connect antiparallel spin-pairs.
\label{fig3}}
\end{figure}

Let us now determine the spin structure enabled by the lattice distortion. 
Previous unpolarized powder and single crystal neutron diffraction measurements \cite{kagomiya07, shlee08} show long range magnetic order with two characteristic wavevectors, ($\frac12$ $\frac12$ 0)$_\textrm{c}$ and (1 0 $\frac12$)$_{\textrm{c}}$. Other powder samples exhibited ($\frac12$ $\frac12$ $\frac12$)$_\textrm{c}$ and (1 0 0)$_\textrm{c}$ reflections as well but the intensity of such peaks varied from sample to sample \cite{shlee08}. Additional work is required to determine whether those reflections are intrinsic or result from strain and/or imperfection. Here we focus on the magnetic structure of samples with only two magnetic wavevectors; ($\frac12$ $\frac12$ 0)$_\textrm{c}$ and (1 0 $\frac12$)$_\textrm{c}$. Single crystal polarized neutron diffraction measurement further provide the important constraint that the ordered moment is confined to the tetragonal basal plane \cite{shlee08}. 
With these preliminaries noted, we determined the full magnetic structure through Rietveld refinement of neutron powder diffraction data. 
In the tetragonal ($I\bar{4}m2$) notation, the wave vectors, ($\frac12$ $\frac12$ 0)$_\textrm{c}$ and (1 0 $\frac12$)$_\textrm{c}$, are equivalent to a single wave vector, $\bm{k}_m$ = (1 0 0)$_\textrm{t}$. 
According to group representation analysis \cite{izyumov}, the magnetic moment at location $\bm{d}_j$ in unit cell $\bm{R}$ is written as $S(\bm{R}, \bm{d}_j) = \sum_{i}C_{i,j}\psi_{i,j}(\bm{d}_j)\textrm{exp}(i\bm{k}_m\cdot\bm{R})$ where $\psi_{i,j}$ is the basis vector of the irreducible representation of the $\bm{k}_m$ subgroup of $I\bar{4}m2$. 
For simplicity, we assume that all spins have equal magnitude. This implies that coefficients $C_{i,j}$ only take on values -1, 0 or 1. 
There are 8 (16) representations for the 8$i$ (16$j$) site that describe structures with spins in the basal plane. 
In total, there are $4(3^8)\times2(3^{16})\simeq2.26\times10^{12}$ coplanar magnetic structures that can be generated by a linear combination of these representations for four 8$i$ and two 16$j$ Cr$^{3+}$ sites in ZnCr$_2$O$_4$. 
When we impose the simplifying and plausible constraints of equal spin magnitudes and zero net moment on each tetrahedron, the number of possible structures is reduced to $\simeq$ $3.5\times10^4$.
When additional constraints are imposed from experimental observations, such as the absence of magnetic reflections at (0 0 $\frac12$)$_\textrm{c}$ = (0 0 1)$_\textrm{t}$ and the intensity ratio $I$((1 1 1)$_\textrm{t}$)/$I$((1 0 2)$_\textrm{t}$) $\sim$ 1 (see Fig. \ref{fig3} (a)), the number is further reduced to 200. 
Comparing these 200 configurations to the neutron powder diffraction data, we found 32 configurations that reproduce the same fit shown as the solid line in Fig. \ref{fig3} (a). 

A common feature of these 32 configurations is that the magnetic moments are along [110]$_\textrm{c}$ directions and every tetrahedron has two pairs of antiparallel spins, (see Fig. \ref{fig3} (b)). 
Furthermore, the non-collinear magnetic structure is orthogonal superposition of two collinear structures with $\bm{k}_1$ = ($\frac12$ $\frac12$ 0)$_\textrm{c}$ and $\bm{k}_2$ = (1 0 $\frac12$)$_\textrm{c}$. 
In the $\bm{k}_1$ structure, there are either three tetrahedra with out-of-plane strong bonds with antiferromagnetic alignments satisfied (yellow squares in Fig. \ref{fig3} (c)) and two tetrahedra with basal plane strong bonds satisfied (yellow crosses in Fig. \ref{fig3} (c)) (type 1) or vice versa (type 1'). 
A $\bm{k}_2$ structure (type 2 in Fig. \ref{fig3} (d)) has four tetrahedra with satisfied out-of-plane bonds and one tetrahedron with satisfied basal plane bonds that matches the strong bond pattern of the blue cell. 
The reverse configuration (type 2') matches the strong bond pattern of the red cell. All 32 configurations involve a combination of these elements. 
Note that in the spin structure (Fig. \ref{fig3} (b)) the pattern of the antiferromagnetic pairs does not exactly match the pattern of the strong NN bonds. 
A single tetrahedron favors a collinear or orthogonal spin structure of two pairs of antiparallel spins if it is made of four strong and two weak bonds or two strong and four weak bonds. 
However, it is impossible to construct such a magnetic structure with the observed $I_{P}$ symmetry \cite{note3} while satisfying the above-mentioned constraints. 

In summary, using single crystal synchrotron X-ray scattering we have determined the tetragonal crystal structure of ZnCr$_2$O$_4$ and we have identified a co-planar spin structure that accounts for the magnetic neutron powder diffraction pattern. Non-satisfied short bond interactions in this structure indicate that isotropic nearest neighbor magneto-elastic interactions alone cannot account for the observed spin-lattice structure. Our experiment therefore calls for a theory of magneto-elastic effects in geometrically frustrated spinel antiferromagnets that includes consideration of Dzyaloshinsky-Moriya and further nearest neighbor exchange interactions.

\begin{acknowledgments}
  We thank O. Tchernyshyov, H. Ueda, M. Gingras for helpful discussions.  Works at Univ. of Virginia and JHU were supported by the U.S. DOE through DE-FG02-07ER45384 and by the U.S. NSF through DMR-0706553, respectively. 
\end{acknowledgments}

{}


\begin{thebibliography}{}
\bibitem{note1} $I\bar{4}c2$ is a subgroup of $Fd\bar{3}m$ with a ($\frac12$ $\frac12$ $\frac12$)$_\textrm{c}$ superlattice wavevector \cite{stokes}. However, $I\bar{4}c2$ is ruled out because we observe (7.5 1.5 1.5)$_\textrm{c}$ and (1.5 1.5 7.5)$_\textrm{c}$  (see Fig. \ref{fig1} (a)) when that space group allows only one of these \cite{ittable}.
\end{thebibliography}
\end{document}